\documentclass[12pt,letterpaper]{article}
\usepackage{osajnl}

\newcommand{\commentout}[1]{}
\commentout{
\documentclass[11pt]{amsart}
\reversemarginpar
\pagestyle{plain}
\setlength{\textwidth}{16.5truecm}
\setlength{\textheight}{22.5truecm}
\setlength{\topmargin}{-0.5truecm}
\setlength{\oddsidemargin}{0cm}
\setlength{\evensidemargin}{\oddsidemargin}

}

\usepackage{amsmath}
\usepackage{amssymb}

\newcommand{\nwc}{\newcommand}

\newcommand{\lt}{\left}
\nwc{\partz}{\frac{\partial }{\partial z}}
\newcommand{\rt}{\right}
\nwc{\ytil}{\tilde{\by}}
\nwc{\al}{\alpha}
\nwc{\half}{\frac{1}{2}}

\newcommand{\lan}{\left\langle}
\newcommand{\ran}{\right\rangle}

\newcommand{\xtil}{{\tilde{\bx}}}

\newcommand{\ks}{{k}}
\newcommand{\bx}{\mathbf x}
\newcommand{\bD}{\mathbf D}
\newcommand{\bp}{\mathbf p}
\newcommand{\br}{\mathbf r}
\newcommand{\bq}{\mathbf q}
\newcommand{\by}{\mathbf y}

\newcommand{\cpv}{\!\!\!\!\!\! - \  }

\nwc{\nwt}{\newtheorem}

\nwt{proposition}{Proposition}
\nwt{lemma}{Lemma}
\nwt{theorem}{Theorem}

\nwt{remark}{Remark}
\nwt{assumption}{Assumption}
\nwt{definition}{Definition} 

\nwc{\bal}{\begin{align}}
\nwc{\be}{\begin{equation}}
\nwc{\ben}{\begin{equation*}}
\nwc{\bea}{\begin{eqnarray}}
\nwc{\beq}{\begin{eqnarray}}
\nwc{\bean}{\begin{eqnarray*}}
\nwc{\beqn}{\begin{eqnarray*}}
\nwc{\beqast}{\begin{eqnarray*}}

\nwc{\eal}{\end{align}}
\nwc{\ee}{\end{equation}}
\nwc{\een}{\end{equation*}}
\nwc{\eea}{\end{eqnarray}}
\nwc{\eeq}{\end{eqnarray}}
\nwc{\eean}{\end{eqnarray*}}
\nwc{\eeqn}{\end{eqnarray*}}
\nwc{\eeqast}{\end{eqnarray*}}

\nwc{\invf}{\cF^{-1}_2}
\nwc{\ep}{\varepsilon}
\nwc{\tep}{\tilde{\varepsilon}}
\nwc{\epsq}{{\varepsilon^2}}
\nwc{\epsqa}{{\varepsilon^{2\alpha}}}
\nwc{\eps}{\varepsilon}
\nwc{\ept}{\epsilon}
\nwc{\vrho}{\varrho}
\nwc{\orho}{\bar\varrho}
\nwc{\ou}{\bar u}
\nwc{\vpsi}{\varpsi}
\nwc{\lamb}{\lambda}

\nwc{\nn}{\nonumber}
\nwc{\bm}{\boldmath}
\nwc{\mf}{\mathbf}
\nwc{\mb}{\mathbf}
\nwc{\ml}{\mathcal}

\nwc{\IA}{\mathbb{A}} 
\nwc{\IB}{\mathbb{B}}
\nwc{\IC}{\mathbb{C}} 
\nwc{\ID}{\mathbb{D}} 
\nwc{\IM}{\mathbb{M}} 
\nwc{\IP}{\mathbb{P}} 
\nwc{\II}{\mathbb{I}} 
\nwc{\IE}{\mathbb{E}} 
\nwc{\IF}{\mathbb{F}} 
\nwc{\IG}{\mathbb{G}} 
\nwc{\IN}{\mathbb{N}} 
\nwc{\IQ}{\mathbb{Q}} 
\nwc{\IR}{\mathbb{R}} 
\nwc{\IT}{\mathbb{T}} 
\nwc{\IZ}{\mathbb{Z}} 

\nwc{\epal}{\ep^{-2\alpha}}
\nwc{\cE}{{\ml E}}
\nwc{\cP}{{\ml P}}
\nwc{\cQ}{{\ml Q}}
\nwc{\cL}{{\ml L}}
\nwc{\cR}{{\ml R}}
\nwc{\cV}{{\ml V}}
\nwc{\cT}{{\ml T}}
\nwc{\crV}{{\ml L}_{(\delta,\rho)}}
\nwc{\cC}{{\ml C}}
\nwc{\cA}{{\ml A}}
\nwc{\cK}{{\ml K}}
\nwc{\cB}{{\ml B}}
\nwc{\cD}{{\ml D}}
\nwc{\cF}{{\ml F}}
\nwc{\cS}{{\ml S}}
\nwc{\cM}{{\ml M}}
\nwc{\cG}{{\ml G}}
\nwc{\cH}{{\ml H}}
\nwc{\bk}{{\mb k}}
\nwc{\cbz}{\overline{\cB}_z}

\nwc{\pft}{\cF^{-1}_\bp}

\newcommand{\fW}{\mathfrak{W}}

\begin{document}

\title{Two-Frequency  Radiative Transfer and
Asymptotic Solution}

\author{Albert C. Fannjiang
 \thanks{
The research is supported in part by the Defense Advanced Research Projects Agency (DARPA) grant 
 N00014-02-1-0603
}
}
\address{Department of Mathematics,
University of California,
Davis, CA 95616-8633}

\email{fannjiang@math.ucdavis.edu} 


\begin{abstract}
Two-frequency radiative transfer (2f-RT) theory is developed for classical
waves in random media. 
Depending on the ratio of the wavelength to the scale
of medium fluctuation 2f-RT equation is either a Boltzmann-like 
integral equation with a complex-valued kernel or a Fokker-Planck-like differential
equation with complex-valued coefficients in the phase space. The 2f-RT equation 
is used to estimate
three physical parameters: the spatial spread,
the coherence length and the coherence bandwidth (Thouless frequency). 
A closed form solution is given for the boundary layer behavior
of geometrical radiative transfer and shows highly nontrivial
dependence of mutual coherence on the spatial displacement
and frequency difference. 
It is shown that the paraxial form of 2f-RT arises naturally
in anisotropic media which  fluctuate slowly
in the longitudinal direction. 
\end{abstract}


\maketitle

\section{Introduction}

Let $U_j, j=1,2,$ be the  random, scalar wave field of
wavenumber
$k_j, j=1,2,$ 
The mutual coherence function and its cross-spectral
version,  known as the two-frequency
mutual coherence function, defined by
\beq
\label{mut}
\Gamma_{12}(\bx,\by)=
\lan U_1(\frac{\bx}{k_1}+\frac{\by}{2k_1})U^*_2(\frac{\bx}{k_2}
-\frac{\by}{2k_2})\ran,
\eeq
where $\lan\cdot\ran$ stands for the ensemble averaging, 
is the central quantity of optical coherence theory,
from which the two-space, two-time correlation function can be
obtained via Fourier transform in frequency,
and therefore plays a
fundamental  role in analyzing propagation of
random pulses  \cite{BW, BM, Ish, MW, SF}. The motivation for the scaling factors in (\ref{mut})
will be given below, cf. (\ref{0.11}).  

In this paper, we set out to analyze the two-frequency mutual coherence as function of  the spatial displacement and
frequency difference for classical waves in 
multiply scattering media. This problem has been extensively
studied in the physics literature (see \cite{BF, Ish, SHG, RN} and references therein). Here we  derive from  the multscale expansion (MSE)
the two-frequency version of the radiative transfer equation 
which is then used to estimate qualitatively the three physical
parameters: the spatial and spatial frequency spreads,
and the coherence bandwidth, also known as the Thouless
frequency in condensed matter physics. Moreover, we show that the boundary layer behavior of the two-frequency
radiative transfer (2f-RT) equation is analytically 
solvable in geometrical optics. The closed form solution (\ref{asym})
provides detailed information
of the two-frequency mutual coherence beyond the current physical
picture \cite{Sha, SHG, RN} (see the discussion about (\ref{current})).  

To this end, we introduce  the two-frequency Wigner
distribution whose ensemble average is equivalent to
the two-frequency mutual coherence
 and 
is a natural extension of
the standard Wigner distribution widely
used in optics \cite{Dra, josaa}.  A different version
of two-frequency Wigner distribution for parabolic waves
was
introduced earlier \cite{2f-whn}
and with it the corresponding radiative transfer equation has been 
derived with full mathematical rigor \cite{2f-crp,2f-rt-physa}. 
In the case of anisotropic media fluctuating slowly in
the longitudinal direction the 2f-RT equation developed here reduces to that of
the paraxial waves in similar media which
lends support to the validity of MSE. 
The other regime where the two frequency radiative transfer
equation has been  obtained  with full mathematical rigor is 
geometrical optics \cite{2f-grt}. 

The main difference between the 2f-RT and the standard theory 
is that the former retains the wave nature of the process 
and is not just about energy transport. Hence the governing equation can not be derived simply based on
the energy conservation law. 

\section{Two-frequency Wigner distribution}

Let $U_j, j=1,2$ be governed  by the reduced wave equation
\beq
\label{helm}
\Delta U_j(\br)+k_j^2 \big(\nu_j+ V_j(\br)\big)U_j(\br)=f_j(\br), \quad\br\in \IR^3, \quad j=1,2
\eeq
where $\nu_j$ and $V_j$ are respectively  the mean 
and fluctuation of the refractive index associated
 with the wavenumber $k_j$ and
are in general complex-valued. The source terms $f_j$ may result
from the initial data or the external sources.
Here and below the vacuum phase speed is set to be
 unity. To solve (\ref{helm}) one needs also
 some boundary condition which is assumed to be
vanishing  at the far field.

We define  the two-frequency Wigner distribution   as
\beq
\label{0.11}
W(\bx,\bp)=\frac{1}{(2\pi)^3}\int
e^{-i\bp\cdot\by}
U_1 (\frac{\bx}{k_1}+
\frac{\by}{2k_1}){U^*_2(\frac{\bx}{k_2}
-\frac{\by}{2k_2})}d\by.
\eeq
In view of  the definition, we see
 that both $\bx$ and $\bp$ are dimensionless.
Here the choice of the scaling factors
 is crucial; namely, the spatial dependence of the wave field should be  measured w.r.t. the probing wavelength. The benefit
 is that this choice leads to a closed form equation for
 $W$. 
 It is easy to see that the ensemble average $\lan W\ran$ is just
 the (partial) Fourier transform of the mutual coherence function
 (\ref{mut}).  
 The two-frequency Wigner distribution defined
 here has a  different scaling factor from the one introduced
 for the parabolic waves  \cite{2f-whn}.
  
 The purpose of introducing the two-frequency Wigner distribution is  to develop a two-frequency
 theory  in analogy to the well studied standard theory of
 radiative transfer. Although the definition  (\ref{0.11}) requires
 the domain to be $\IR^3$,  the governing  radiative transfer equation, 
 once obtained,  can be (inverse) Fourier transformed back to get the governing equation 
 for the two-point function $U_1(\br_1)U_2^*(\br_2)$ or $\Gamma_{12}$ as their boundary conditions
 are usually easier to describe (cf. eq. (\ref{mean-eq2})).  
 
 The Wigner distribution 
has  the following easy-to-check properties:
 \beq
\int |W|^2(\bx,\bp)d\bx d\bp&=&
\lt(\frac{\sqrt{\ks_1\ks_2}}{2\pi}\rt)^{3}
\int |U_1|^2(\bx)d\bx \int |U_2|^2(\bx)d\bx\nn\\
\label{2.2.2}
\int W(\bx,\bp)e^{i\bp\cdot\by}d\bp&=&
U_1
(\frac{\bx}{k_1}+
\frac{\by}{2k_1})
U_2^{*}(\frac{\bx}{
k_2}
-\frac{\by}{2k_2})\\
\int W(\bx,\bp)e^{-i\bx\cdot
\bq}d\bx&=&\lt({\pi^2k_1k_2}\rt)^{3}
\widehat U_1(\frac{k_1\bp}{4}
+\frac{k_1\bq}{2})
{\widehat U}^{*}_2(\frac{k_2\bp}{4}
-\frac{k_2\bq}{2}),
\eeq
where $\widehat{\cdot}$ stands for the Fourier transform, 
and hence contains all the information 
in the two-point two-frequency function. In particular,
\beqn
\int \bp W(\bx,\bp)d\bp&=&-i\Big[\frac{1}{2k_1}
\nabla U_1(\frac{\bx}{k_1}) U_2^*(\frac{\bx}{k_2})-\frac{1}{2k_2}
U_1(\frac{\bx}{k_1}) \nabla U_2^*(\frac{\bx}{k_2})
\Big]
\eeqn
which, in the case of $k_1=k_2$, is proportional to
the energy flux density. 

We now derive the equation for the two-frequency
Wigner distribution.  
After taking the derivative $\bp\cdot\nabla$ and
some calculation we have
\beq
\nn
\bp\cdot\nabla W&=&\frac{i}{2(2\pi)^3}
\int e^{-i\bp\cdot\by} U_1 (\frac{\bx}{k_1}+
\frac{\by}{2k_1}){U^*_2(\frac{\bx}{k_2}
-\frac{\by}{2k_2})}V_1(\frac{\bx}{k_1}+
\frac{\by}{2k_1})d\by\\
&&
-\frac{i}{2(2\pi)^3}
\int e^{-i\bp\cdot\by} U_1 (\frac{\bx}{k_1}+
\frac{\by}{2k_1}){U^*_2(\frac{\bx}{k_2}
-\frac{\by}{2k_2})}V_2^*(\frac{\bx}{k_2}-
\frac{\by}{2k_2})d\by\nn\\
&&+\frac{i}{2}(\nu_1-\nu^*_2) W+F \label{raw}
\eeq
where the function $F$ depends linearly
on $U_j$ and $f_j$:
\beq
F&=&-\frac{i}{2(2\pi)^3}
\int e^{-i\bp\cdot\by} f_1 (\frac{\bx}{k_1}+
\frac{\by}{2k_1}){U^*_2(\frac{\bx}{k_2}
-\frac{\by}{2k_2})}d\by\nn\\
&&
+\frac{i}{2(2\pi)^3}
\int e^{-i\bp\cdot\by} U_1 (\frac{\bx}{k_1}+
\frac{\by}{2k_1}){f^*_2(\frac{\bx}{k_2}
-\frac{\by}{2k_2})}d\by.\label{fcn}
\eeq
Substituting the spectral representation of $V_j$
\beq
\label{spec}
V_j(\bx)=\int e^{i\bq\cdot\bx} \hat V_j(d\bq)
\eeq
in the expression and using the definition of $W$ we then
obtain the exact equation
\beq
\label{WME}
\lefteqn{\bp\cdot\nabla W-\frac{i}{2}(\nu_1-\nu^*_2) W-F}\\
&=&
\frac{i}{2}\int \hat V_1(d\bq)
e^{i\bq\cdot\bx/k_1} W(\bx,\bp-\frac{\bq}{2k_1})-
\frac{i}{2}\int \hat V^*_2(d\bq)
e^{-i\bq\cdot\bx/k_2} W(\bx,\bp-\frac{\bq}{2k_2}).\nn
\eeq
Here and below $\hat V_2^*$ is the complex-conjugate
of the Fourier spectral measure $\hat V_2$. The full derivation of (\ref{WME}) is given in
Appendix A.

Let us pause to compare the classical wave with
the quantum wave function in the context of two-frequency
formulation.
The quantum wave functions $\Psi_j$ at two
different frequencies $\omega_1,\omega_2$ satisfy
the stationary Schr\"odigner equaiton 
\beq
\label{sch2}
\frac{\hbar^2}{2}\Delta \Psi_j+ \big(\nu_j+ V_j(\bx)\big)\Psi_j&=&
-\omega_j \hbar \Psi_j+f_j,\quad j=1,2,
\eeq
where $\nu_j+V_j$ are  hypothetical, energy-dependent
real-valued potentials. Here the source terms $f_j$ equal the initial data $f$ of the time dependent problem. Usually in the quantum  mechanical
context, the potential function does not explicitly depend on
the energy level (i.e. dispersionless). 

The natural definition of the two-frequency Wigner distribution 
for the quantum wave functions is 
\beq
W(\bx,\bp)=\frac{1}{(2\pi)^3}
\int e^{-i\bp\cdot\by} \Psi_1(\bx+\frac{\hbar\by}{2})
\Psi^*_2(\bx-\frac{\hbar\by}{2})d\by
\eeq
which satisfies the Wigner-Moyal equation
\beq
\label{WM2}
\lefteqn{\bp\cdot\nabla W+i(\omega_2-\omega_1)W+\frac{i}{\hbar}(\nu_2^*-\nu_1)W }\\
&=&\frac{i}{\hbar}\int \hat V_1(d\bq)
e^{i\bq\cdot\bx} W(\bx,\bp-\frac{\hbar\bq}{2})-
\frac{i}{\hbar}\int \hat V^*_2(d\bq)
e^{-i\bq\cdot\bx} W(\bx,\bp-\frac{\hbar\bq}{2})+F\nn
\eeq
where $F$ has a similar expression to (\ref{fcn}). 
The main difference between the quantum and classical waves
in the Wigner formulation is that the derivation of a closed-form  
equation does not require
rescaling  each energy component w.r.t. its de Broglie
wavelength. The implication in radiative transfer  will be further discussed
 (see the remark following eq. (\ref{rt-sch})).

\section{Two-frequency radiative transfer scaling}
We assume that $V_j(\bx), j=1,2$ are {\em real-valued}, centered, random stationary
(i.e. 
statistically homogeneous) {\em  ergodic}  field
admitting the spectral representation
(\ref{spec}) with the spectral measures $\hat{V}_j(d\bp), j=1,2$
such that
\[
\lan \hat{V}_j(d\bp)\hat{V}_j^*(d\bq)\ran
=\delta(\bp-\bq)\Phi_j(\bp)d\bp d\bq
\]
\commentout{
$\hat{V}_j(d\bp)=A_j(d\bp)+iB_j(d\bp), j=1,2,$ where
$A_j$ and $B_j$ are the real and imaginary parts, respectively. 
We assume that for $j=1,2$
\beq
\lan A_j(d\bp)A_j(d\bq)\ran=\lan B_j(d\bp)B_j(d\bq)\ran
&=&\frac{1}{2}\delta(\bp-\bq)\Phi_j(\bp)d\bp d\bq,\\
\lan A_j(d\bp)B_j(d\bq)\ran&=&0, \quad\forall \bp, \bq
\eeq
}
where $\Phi_j$ are the (nonnegative-valued) power spectral densities of
the random fields $V_j, j=1,2$. The above $\delta$ function
is a consequence of  the statistical homogeneity of
the random field $V_j$. 
As $V_j, j=1,2$ are real-valued, $\hat V^*_j (d\bp)=\hat V_j(-d\bp)$ and hence
the power spectral densities $\Phi_j(\bp)$
satisfy the symmetry property $\Phi_j(\bp)=\Phi_j(-\bp),\forall \bp$.

We will also need the cross-frequency correlation
and we postulate the existence of
the cross-frequency spectrum $\Phi_{12}$ such
that 
\[
\lan \hat{V}_1(d\bp)\hat{V}_2^*(d\bq)\ran=
\delta(\bp-\bq)\Phi_{12}(\bp)d\bp d\bq.
\]
Here $\Phi_{12}$ needs not be real-valued. 
\commentout{
\beq
\lan A_1(d\bp)A_2(d\bq)\ran=\lan B_1(d\bp)B_2(d\bq)\ran
&=&\frac{1}{2}\delta(\bp-\bq)\Phi_{12}(\bp)d\bp d\bq,\\
\lan A_1(d\bp)B_2(d\bq)\ran=\lan A_2(d\bp)B_1(d\bq)\ran
&=&0, \quad\forall \bp, \bq
\eeq
}

An important regime of multiple scattering of classical waves
takes
place when the scale of medium fluctuation  is much smaller than
the propagation distance but is comparable or
much larger than the  wavelength \cite{Ish, Mis}.
Radiative transfer regime can be characterized by the scaling limit
which replaces $\nu_j+V_j$ in eq. (\ref{helm}) with
\beq
\label{scaling}
\frac{1}{\theta^2\ep^2}\Big(\nu_j+ \sqrt{\ep}V_j(\frac{\br}{\ep})
\Big),\quad \theta>0,\quad\ep\ll 1
\eeq
 where $\ep$ is 
the ratio of the scale of medium fluctuation to the $O(1)$ propagation distance and $\theta$ the ratio
of the wavelength to the scale of medium fluctuation. 
Hence $\theta\ep$ is the ratio of the wavelength
to the propagation distance and 
the prefactor $(\theta\ep)^{-2}$ arises from 
rescaling the wavenumber 
$k\to k/(\ep\theta)$. This is so called the
weak coupling (or disorder) limit in kinetic theory which prohibits 
 the Anderson localization from happening \cite{Spo}. 
Note that the  resulting
 medium fluctuation 
 $
 {\ep^{-3/2}} V_j({\br}/{\ep})
 $
converges to a spatial white-noise in three dimensions. 
 
Physically  speaking the radiative transfer scaling belongs to   the diffusive wave  regime under the condition of a large dimensionless conductance $g= N\ell_t/L$,
where $\ell_t$ is the transport mean free path,
 $L$ is the sample size in the direction of propagation
 and $N=2\pi A/\lambda^2$ is the number of transverse modes,  limited by the illuminated area $A$
and the wavelength of radiation $\lambda$ \cite{BF, SHG}.
The dimensionless conductance $g$ can be expressed as $g=k\ell_t/\hbox{Fr}$
with the inverse Fresnel number $\hbox{Fr}=\lambda L/A$. 
With the scaling (\ref{scaling}), $k\ell_t \sim \hbox{Fr}^{-1}\sim \theta^{-1}\ep^{-1}$
 and hence
$g\sim \theta^{-2}\ep^{-2}\gg 1$ for any finite $\theta$ as $\ep\to 0$.

Anticipating small-scale fluctuation due to (\ref{scaling})  we modify the definition of
the two-frequency Wigner distribution in
the following way
\beqn
W(\bx,\bp)=\frac{1}{(2\pi)^3}\int
e^{-i\bp\cdot\by}
U_1 (\frac{\bx}{k_1}+
\frac{\theta\ep\by}{2k_1}){U^*_2(\frac{\bx}{k_2}
-\frac{\theta\ep\by}{2k_2})}d\by
\eeqn
 Eq. (\ref{WME}) now becomes
\beq
\label{wig}
{\bp\cdot\nabla W}-F
&=&\frac{i}{2\ep\theta}(\nu_1-\nu^*_2) W+\frac{1}{\sqrt{\ep}}\cL W
\eeq
where the operator $\cL$ is defined by
\beq
\cL W(\bx,\bp)&=&
\frac{i}{2\theta}\int \hat V_1(d\bq)
e^{i\frac{\bq\cdot\bx}{\ep k_1}} W(\bx,\bp-\frac{\theta\bq}{2k_1})-\frac{i}{2\theta}\int \hat V^*_2(d\bq)
e^{-i\frac{\bq\cdot\bx}{\ep k_2}} W(\bx,\bp-\frac{\theta\bq}{2k_2}).\nn
\eeq
To capture the cross-frequency correlation
in the radiative transfer regime  we also need 
to restrict the frequency difference range 
\beq
\label{band}
\lim_{\ep\to 0}\ks_1=\lim_{\ep\to 0}\ks_2=\ks,
\quad \frac{\ks_2-\ks_1}{\ep\theta k}=\beta
\eeq
where $k, \beta>0$ are independent of $\ep$ and $\theta$. 
Assuming the differentiability  of the mean refractive index's dependence
on the wavenumber we write
\beq
\label{band'}
\frac{\nu_2^*-\nu_1}{2\ep\theta}=\nu'
\eeq
where $\nu'$ is independent of $\ep, \theta$. 

\section{Multi-scale expansion (MSE)}
\label{sec-mse}
To derive the radiative transfer equation for
the two-frequency Wigner distribution we employ
MSE \cite{BLP, RPK} which  begins with introducing the fast variable
\[
\xtil=\bx/\ep
\]
and treating $\xtil$ as independent from the slow variable
$\bx$. Consequently the derivative $\bp\cdot\nabla$ consists of two terms
\beq
\label{fast}
\bp\cdot\nabla=\bp\cdot\nabla_\bx+\ep^{-1}\bp\cdot\nabla_{\xtil}.
\eeq
Then MSE posits  the following asymptotic expansion:
\beq
\label{mse}
W(\bx,\bp)=\bar W(\bx,\xtil,\bp)+\sqrt{\ep} W_1(\bx,\xtil,\bp)+\ep W_2(\bx,\xtil,\bp)+O(\ep^{3/2}),\quad \xtil=\bx\ep^{-1}
\eeq
whose proper sense will be explained 
below. 

Substituting the ansatz into eq. (\ref{wig}) and using 
(\ref{fast}) we determine each term of (\ref{mse})
by equating terms of the same order of magnitude
starting with the highest order $\ep^{-1}$.

The $\ep^{-1}$-order  equation has one term:
\beqn
\bp\cdot\nabla_{\xtil} \bar W=0
\eeqn
which can be solved by setting $\bar W=\bar W(\bx,\bp)$.
Namely, to the leading order $W$ is independent of
the fast variable. Since the fast variable is due to
medium fluctuation, this suggests that $\bar W$ is
deterministic. 

The next is the $\ep^{-1/2}$-order equation:
\beq
\label{w1}
\bp\cdot\nabla_{\xtil} W_1=\cL\bar W.
\eeq
\commentout{
\frac{i}{2}\int \hat V_1(d\bq)
{e^{i\frac{\bq\cdot\xtil}{k_1}}}\bar W(\bx,\bp-\frac{\bq}{2k_1})-
\frac{i}{2}\int \hat V^*_2(d\bq)
e^{-i\frac{\bq\cdot\xtil}{ k_2}} \bar W(\bx,\bp-\frac{\bq}{2k_2})
\eeqn
}
We seek a solution that is stationary in $\xtil$, square-integrable in $\bp$ and
has finite second moment. The solvability condition
(Fredholm alternative) 
is that the right hand side, $\cL \bar W$, satisfies $\int\IE\big[
\Psi^* \cL\bar W\big]d\bp=0$
for any $\xtil$-stationary, square-integrable field $\Psi(\xtil,\bp)$ satisfying $\bp\cdot\nabla_\xtil \Psi=0$. 
The solvability condition is, however, not easy to enforce. 
Alternatively we consider the regularized equation 
\beq
\label{w12}
\ep W^\ep_1+\bp\cdot\nabla_\xtil W^\ep_1=
\cL\bar W
\eeq
which is always solvable for $\ep>0$ and admits the solution  
\beq
\label{w1'}
W^\ep_1(\bx,\xtil,\bp)&=&
\frac{i}{2\theta}\int \hat V_1(d\bq)
\frac{e^{i\frac{\bq\cdot\xtil}{k_1}}}{\ep+i\bq\cdot\bp/k_1} \bar W(\bx,\bp-\frac{\theta\bq}{2\ks_1})\\
&&-
\frac{i}{2\theta}\int \hat V^*_2(d\bq)
\frac{e^{-i\frac{\bq\cdot\xtil}{k_2}}}{\ep-i\bq\cdot\bp/k_2} \bar W(\bx,\bp-\frac{\theta\bq}{2\ks_2}).\nn
\eeq
In the jargons of asymptotic analysis \cite{BLP}, $\sqrt{\ep} W_1^\ep$ is
called the first {\em corrector}. 
In order to control  the first corrector,
we choose $\bar W$ such that $\cL\bar W$ has zero mean. 
This is a necessary condition as we seek a $\xtil$-stationary solution and consequently $\lan \bp\cdot\nabla_{\tilde\bx} W_1\ran=\bp\cdot\nabla_{\xtil}\lan W_1\ran=0$. Needless to say, this condition
is weaker than the solvability condition stated above
and is satisfied  for any {\em deterministic} $\bar W$ since both
$V_1$ and $V_2$ have zero mean. 

Indeed, under the assumption of deterministic $\bar W$,
the resulting equation will be much simplified so we impose
this property on $\bar W$ from now on.  The fact that in the limit $\bar W$
is deterministic  can be proved rigorously in the paraxial regime \cite{2f-rt-physa}. 

Finally the $O(1)$ equation is
\beq
\nn
\bp\cdot\nabla_\xtil W_2(\bx,\xtil, \bp)&=& -\bp\cdot\nabla_\bx \bar W(\bx,\bp)
-i\nu'\bar W + F
+\frac{i}{2\theta}\int \hat V_1(d\bq)
{e^{i\frac{\bq\cdot\xtil}{k_1}}}W^\ep_1(\bx, \xtil, \bp-\frac{\theta\bq}{2\ks_1})\\
&&-
\frac{i}{2\theta}\int \hat V^*_2(d\bq)
e^{-i\frac{\bq\cdot\xtil}{ k_2}} W^\ep_1(\bx,\xtil, \bp-\frac{\theta\bq}{2\ks_2})\label{w2}
\eeq
which can be solved with regularization  as in (\ref{w12}) and yields
the second corrector ${\ep}W^\ep_2$.
Again we impose on the right hand side of (\ref{w2}) the weaker
condition of zero mean. 
 Using (\ref{w1'}) in (\ref{w2}),
 taking the ensemble average and passing
 to the limit $\ep\to 0$
 we obtain the governing equation
for $\bar W$: 
\beqn
\lefteqn{\bp\cdot\nabla_\bx  \bar W(\bx,\bp)+i\nu' \bar W-\lan F\ran }\\
&=&-\frac{k_1^{3}}{2\theta^{4}}
\int  d\bq \Phi_1\big(\frac{k_1}{\theta}(\bp-\bq)\big)\pi\delta(|\bp|^2
-{|\bq|^2}) \bar W (\bx,\bp)+\frac{ik_1^{3}}{2\theta^{4}}
\int\cpv d\bq\frac{\Phi_1\big(\frac{k_1}{\theta}(\bp-\bq)\big)}{|\bp|^2-|\bq|^2}
\bar W (\bx,\bp)\\
&&\nn-\frac{k_2^{3}}{2\theta^{4}} 
\int  d\bq \Phi_2\big(\frac{k_2}{\theta}(\bp-\bq)\big)\pi\delta(|\bp|^2
-{|\bq|^2})\bar W (\bx,\bp)-\frac{ik_2^{3}}{2\theta^{4}}
\int\cpv d\bq\frac{\Phi_2\big(\frac{k_2}{\theta}(\bp-\bq)\big)}{|\bp|^2-|\bq|^2}
\bar W (\bx,\bp)\\
&&+\frac{1}{4\theta^2}
\int d\bq \Phi_{12}(\bq)e^{i\xtil\cdot\bq (k_1^{-1}-k^{-1}_2)}\pi\delta\big(\frac{\bq}{k_2}\cdot(\bp-\frac{\theta\bq}{2k_1})\big) \bar W (\bx, \bp-\frac{\theta\bq}{2k_1}-\frac{\theta\bq}{2k_2})\\
&&\nn
+\frac{1}{4\theta^2}
\int d\bq \Phi_{12}(\bq)e^{i\xtil\cdot\bq(k_1^{-1}-k^{-1}_2)}\pi\delta\big(\frac{\bq}{k_1}\cdot(\bp-\frac{\theta\bq}{2k_2})\big)\bar W (\bx, \bp-\frac{\theta\bq}{2k_1}-\frac{\theta\bq}{2k_2})\\
&&
+\frac{i}{4\theta^2}\int\cpv d\bq\Big[\frac{1}{\frac{\bq}{k_2}\cdot(\bp-\frac{\theta\bq}{2k_1})}
-\frac{1}{\frac{\bq}{k_1}\cdot(\bp-\frac{\theta\bq}{2k_2})}\Big]
\Phi_{12}(\bq)e^{i\tilde\bx\cdot\bq(k_1^{-1}-k^{-1}_2)}\bar W(\bx, \bp-\frac{\theta\bq}{2k_1}-\frac{\theta\bq}{2k_2})
\commentout{
+\frac{i}{4\theta^2}\int\cpv\frac{\Phi_{12}(\bq)e^{i\bx\cdot\bq\beta\theta}}{\frac{\bq}{k_2}\cdot(\bp-\frac{\theta\bq}{2k_1})}
\bar W(\bx, \bp-\frac{\theta\bq}{2k_1}-\frac{\theta\bq}{2k_2})
-\frac{i}{4\theta^2}\int\cpv\frac{\Phi_{12}(\bq)e^{i\bx\cdot\bq\beta\theta}}{\frac{\bq}{k_1}\cdot(\bp-\frac{\theta\bq}{2k_2})}
\bar W (\bx, \bp-\frac{\theta\bq}{2k_1}-\frac{\theta\bq}{2k_2})
}
\eeqn
where we have used the fact that in the sense of generalized function 
\[
\lim_{\eta\to 0} \frac{1}{\eta+i\xi}= \pi \delta(\xi)-\frac{i}{\xi}
\]
with  the second term giving rise to the  Cauchy principal value integral denoted by $\int\cpv$. From (\ref{fcn}) we have
the expression for $\lan F\ran$
\beqn
\lan F\ran &=&-\frac{i}{2(2\pi)^3}
\int e^{-i\bp\cdot\by} f_1 (\frac{\bx}{k_1}+
\frac{\by}{2k_1})\lan {U^*_2(\frac{\bx}{k_2}
-\frac{\by}{2k_2})}\ran d\by\nn\\
&&
+\frac{i}{2(2\pi)^3}
\int e^{-i\bp\cdot\by}\lan  U_1 (\frac{\bx}{k_1}+
\frac{\by}{2k_1})\ran {f^*_2(\frac{\bx}{k_2}
-\frac{\by}{2k_2})}d\by.
\eeqn
which  depends only on the mean fields $\lan U_1\ran,
\lan U_2\ran$, both assumed known throughout the paper. 

Putting all the terms together with the regularization 
we arrive at the following MSE
\beq
\label{mse22}
W(\bx,\bp)=\bar W(\bx,\bp)+\sqrt{\ep} W^\ep_1(\bx,\xtil,\bp)+\ep W^\ep_2(\bx,\xtil,\bp)
\eeq
which satisfies 
\beq
\label{mse2}
\lefteqn{\Big(\bp\cdot\nabla -\frac{1}{\sqrt{\ep}} \cL\Big)W+i\nu' W-F}\\
&=&
(i\nu'-1)\sqrt{\ep} W^\ep_1 +\sqrt{\ep} \bp\cdot\nabla_\bx W^\ep_1-\sqrt{\ep} \cL W^\ep_2 +(i\nu'-1)\ep W_2^\ep
 +\ep \bp\cdot\nabla_\bx W^\ep_2.\nn 
\eeq
Unfortunately the right hand side of (\ref{mse2}) 
does not vanish in the strong $L^2$-topology  but only in the weak topology as in 
\be
\label{corr22}
\lim_{\ep\to
0}\ep \int\,d\bx\,\lan\left|\int\,d\bp\,W^\ep_1(\bx,\frac{\bx}{\ep},\bp)\psi(\bp)\right|^2\ran
=0,\quad \forall \psi \in L^2
\ee
(see Appendix B). It is not clear at this point how to
 justify the preceding argument and construction of asymptotic solution with full mathematical rigor. Fortunately, in the regime of geometrical optics, the rigorous asymptotic result
can be obtained  by a probabilistic method \cite{2f-grt}
and is the same as derived by MSE
(see Section \ref{sec-grt}). Another regime for
which the asymptotic result 
can be fully justified is paraxial waves which we will
turn to in the next section. 

Due to the assumption (\ref{band}) and the assumed continuous dependence of the medium fluctuation on the frequency
 we have $\lim\Phi_1=\lim\Phi_2=\lim\Phi_{12}=\Phi$.
 As a consequence,  all the Cauchy principal value integrals  cancel out. With some
changes of variables  
the governing equation for $ \bar W$ 
takes  the much simplified form:
\beq
\label{wb}
\lefteqn{\bp\cdot\nabla_\bx  \bar W+ i\nu' \bar W-\lan F\ran}\\
&=&\frac{\pi k^{3}}{\theta^{4}}
\int  d\bq \Phi\big(\frac{k}{\theta}(\bp-\bq)\big)\delta(|\bp|^2
-{|\bq|^2})\Big[e^{i\bx\cdot(\bp-\bq )\beta} 
\bar W \big(\bx,\bq\big)
-\bar W (\bx,\bp)\Big].\nn
\eeq
The $\delta$-function in the scattering kernel is 
due to elastic scattering which preserve the
wavenumber. 
 When $\beta=0$ (then $\nu_1=\nu_2$ and  $i\nu' \sim $  the imaginary part of $\nu$), eq. (\ref{wb}) 
reduce to the standard form of radiative transfer equation 
for the phase space energy density \cite{Sch, Hop, Cha, Mis}.  For
$\beta>0$,  the wave featue is retained in (\ref{wb}). When $\beta\to\infty$,
the first term in the bracket on the right hand side of (\ref{wb}) drops out,
due to rapid phase fluctuation,  so the random scattering effect 
  is pure damping:
  \beq
\label{damp}
{\bp\cdot\nabla_\bx \bar W + i\nu' \bar W-\lan F\ran}
&=&-\frac{\pi k^{3}}{\theta^{4}}
\int  d\bq \Phi\big(\frac{k}{\theta}(\bp-\bq)\big)\delta(|\bp|^2
-{|\bq|^2}) \bar W (\bx,\bp).\nn
\eeq

As a comparison, for Schr\"odinger equation (\ref{sch2}) 
in the frequency domain,
we modify the Wigner distribution as
\beqn
W(\bx,\bp)=\frac{1}{(2\pi)^3}
\int e^{-i\bp\cdot\by} \psi_1(\bx+\frac{\ep\hbar\by}{2})
\psi^*_2(\bx-\frac{\ep\hbar\by}{2})d\by
\eeqn
and in the limit $\ep\to 0$ obtain the radiative transfer
equation following the same procedure
\beq
\label{rt-sch}
\lefteqn{\bp\cdot\nabla_\bx \bar W+ i(\omega_2-\omega_1)\bar W+\frac{2i}{\hbar}\nu' \bar W-\lan F\ran}\\
&=&\frac{4\pi }{\hbar^{4}}
\int  d\bq \Phi\big(\frac{\bp-\bq}{\hbar}\big)\delta(|\bp|^2
-{|\bq|^2})\Big[
\bar W \big(\bx,\bq\big)
-\bar W(\bx,\bp)\Big].\nn
\eeq
The absence  of the factor $e^{i\bx\cdot(\bp-\bq )\beta} $ in
eq. (\ref{rt-sch}), and therefore  the cross-frequency interference, is the main characteristic of
2f-RT for 
quantum waves.

\commentout{
The convergence of
the above  scaling limit is probably  provable 
by extending the rigorous diagrammatic method 
developed for the time dependent Schr\"odinger
equation in \cite{Sp}, \cite{Sp2}, \cite{HLW}, \cite{EY}. 
Here instead of the time dependent Schr\"odinger equation,
we have the stationary Schr\"odinger equation with two
different energy-dependent potentials.  
However, the diagrammatic approach, rigorous or not,  is more complicated
to carry out 
and we will be content  to give an explanation 
in line with the multi-scale expansion in  Appendix B. 
}
\section{Paraxial 2f-RT: anisotropic medium}
\label{prt}

Forward-scattering approximation, also called paraxial approximation,  is valid when  back-scattering is negligible
and, as we show now, this is the case for anisotropic media fluctuating 
slowly  in the (longitudinal) direction of propagation.  Let  $z$ denote the longitudinal coordinate and $\bx_\perp$ the transverse coordinates. Let $p$ and $\bp_\perp$ denote the longitudinal and
transverse components of $\bp\in \IR^3$, respectively. 
Let  $\bq=(q, \bq_\perp)\in \IR^3$ be likewise defined.

Consider now  a highly anisotropic spectral density
for  a medium fluctuating much more
slowly in the longitudinal direction, i.e.
replacing $\Phi\big((\bp-\bq)k/\theta\big)$ in (\ref{wb}) by 
\[
\frac{1}{\eta}\Phi\lt(\frac{k}{\eta \theta}(p-q), \frac{k}{\theta} (\bp_\perp-\bq_\perp)\rt),\quad\eta\ll 1,
\]
which, in the limit $\eta\to 0$, tends to
\beq
\label{aniso}
\frac{\theta}{k}\delta(p-q) \int dw \Phi\lt(w, 
\frac{k}{\theta} (\bp_\perp-\bq_\perp)\rt). 
\eeq
Writing 
$ \bar W= \bar W(z,\bx_\perp,p, \bp_\perp)$
we can  approximate    eq. (\ref{wb}) by \beq
\nn
\lefteqn{p\partial_z \bar W+\bp_\perp\cdot\nabla_{\bx_\perp}\bar W+i\nu' \bar W-\lan F\ran}\\
&=&\frac{\pi k^{2}}{\theta^{3}}
\int  d\bq_\perp \int dw \Phi\big(w, \frac{k}{\theta}(\bp_\perp-\bq_\perp)\big)\delta(|\bp_\perp|^2
-{|\bq_\perp|^2})\nn \\
&&\times \Big[e^{i\bx_\perp\cdot(\bp_\perp-\bq_\perp)\beta} 
\bar W\big(z, \bx_\perp,p, \bq_\perp\big)
-\bar W(z, \bx_\perp,p, \bp_\perp)\Big].\label{rt-para}
\eeq
Eq. (\ref{rt-para}) is identical to  the 2f-RT equation
rigorously derived directly 
from the paraxial wave equation for similar
 anisotropic media \cite{2f-crp, 2f-rt-physa}. This is somewhat surprising in view
of the different scaling factors in the definition
of two-frequency Wigner distributions in the two cases. 

Note that in eq. (\ref{rt-para})
the longitudinal momentum $p$ plays the role
of a parameter and does not change during propagation and scattering. An important implication of this observation is
that eq. (\ref{rt-para}) can be solved as an evolution equation in
the direction of increasing $z$ with the {\em one-sided} boundary condition (e.g. at $z=\hbox{const.}$).  
In other words, the influence from the other boundary
vanishes as the longitudinal direction is infinitely long. 
The initial value problem  of (\ref{rt-para}) is much
easier to solve than the boundary value problem of (\ref{wb}).

\section{Two-frequency geometrical radiative transfer (2f-GRT)}
\label{sec-grt}
Let us consider the further limit $\theta\ll 1$ when the wavelength is much shorter
than the correlation length of the medium fluctuation. To this end, the following form 
is more convenient to work with
\beq
\label{wb2}
\lefteqn{\bp\cdot\nabla_\bx \bar W + i\nu' \bar W-\lan F\ran }\\
&=&\frac{\pi k}{2\theta^2}
\int  d\bq \Phi\big(\bq \big)\delta\big(\bq\cdot(\bp-\frac{\theta\bq}{2k})\big)\Big[e^{i\bx\cdot\bq \beta\theta/k} 
\bar W\big(\bx,\bp-\frac{\theta\bq}{k}\big)
-\bar W(\bx,\bp)\Big]\nn
\eeq
which is obtained from eq. (\ref{wb}) after a change of
variables. 
We expand the right hand side of (\ref{wb2}) in $\theta$  and pass to the limit
$\theta\to 0$ to obtain
\beq
\label{go}\label{fp}
{\bp\cdot\nabla_\bx \bar W+ i\nu' \bar W-\lan F\ran }
&=&
\frac{1}{4k}\lt(\nabla_\bp-i{\beta}\bx\rt)
\cdot \bD\cdot 
\lt(\nabla_\bp-i{\beta}\bx\rt)  \bar W
\eeq
with the (momentum) diffusion coefficient
\beq
\label{diffusion}
\bD(\bp)=\pi\int \Phi(\bq)\delta(\bp\cdot\bq)\bq\otimes\bq d\bq.
\eeq
The symmetry $\Phi(\bp)=\Phi(-\bp)$ plays an explicit role here in rendering the right hand side of eq. (\ref{wb2})
a second-order operator in the limit $\theta\to 0$. 
Eq. (\ref{go}) can be rigorously derived
from geometrical optics  by a probabilistic method
\cite{2f-grt}. 

\subsection{Spatial (frequency) spread and coherence bandwidth}
\label{sec-iso}
Through dimensional analysis, eq. (\ref{go}) 
yields qualitative information about 
important physical parameters of the stochastic medium.
To show this, let us  assume for simplicity  the isotropy of the medium, i.e. $\Phi(\bp)=\Phi(|\bp|)$,
 so that $\bD={C} |\bp|^{-1} \Pi(\bp)$ where 
 \beq
 \label{const}
{C}=\frac{\pi}{3}\int\delta\Big(\frac{\bp}{|\bp|}\cdot\frac{\bq}{|\bq|}\Big)\Phi(|\bq|)|\bq| d\bq
\eeq
is a constant
 and $\Pi(\bp)$ the orthogonal projection
onto the plane perpendicular to $\bp$. 
In view of (\ref{go}) $C$ (and $\bD$)  has the dimension of inverse length while the variables $\bx$ and $\bp$ are dimensionless. 

Now consider the following change of variables
\beq
\label{change}
\bx=\sigma_x k\tilde \bx,\quad \bp=\sigma_p\tilde\bp/k,
\quad \beta=\beta_c\tilde\beta
\eeq
where $\sigma_x$ and $\sigma_p$ are respectively the
spreads in position  and spatial frequency, and $\beta_c$ is
the coherence bandwidth. Let us substitute (\ref{change})
into eq. (\ref{fp}) and aim for the standard form
\beq
\label{stan}
{\tilde\bp\cdot\nabla_{\tilde\bx} \bar W  + i\nu' \bar W-\lan F\ran }
&=&
\lt(\nabla_{\tilde\bp}-i{\tilde\beta}\tilde\bx\rt)
\cdot |\tilde\bp|^{-1}\Pi(\tilde\bp)
\lt(\nabla_{\tilde\bp}-i{\tilde\beta}\tilde\bx\rt)  \bar W.
\eeq
The 1-st term on the left side yields the first duality relation
\beq
\label{du1}
\sigma_x/\sigma_p\sim 1/k^2.
\eeq
The balance of terms in each pair of parentheses yields
the second duality relation
\beq
\label{du2}
\sigma_x\sigma_p\sim \frac{1}{\beta_c}
\eeq
whose left hand side is the {\em space-spread-bandwidth product.}
Finally the removal of the constant $C$ determines
\beq
\sigma_p\sim k^{2/3} C^{1/3}
\eeq
from which 
 $\sigma_x$ and $\beta_c$ can be determined
 by using (\ref{du1}) and (\ref{du2}):
 \[
 \sigma_x\sim k^{-4/3} C^{1/3},\quad \beta_c\sim k^{2/3} C^{-2/3}.
 \]

We do not know if, as it stands, eq. (\ref{stan}) is analytically solvable
but we can solve analytically for its boundary layer behavior.

\subsection{Boundary layer asymptotics: paraxial 2f-GRT}
Consider the half space $z\geq 0$ occupied by the random medium  and  a collimated narrow-band beam propagating
in the $z$ direction and incident normal to
the boundary ($z=0$) of the medium. Near the point of incidence on the boundary the corresponding two-frequency Wigner distribution 
would be highly concentrated at the longitudinal momentum,
say, 
$p=1$. Hence  we can assume that the projection $\Pi(\bp)$ in
(\ref{stan}) is effectively just the projection onto the transverse plane coordinated by $\bx_\perp$
and approximate eq. (\ref{go}) by 
\beq
\label{para}
{\Big[\partial_{z}+{\bp_\perp\cdot\nabla_{\bx_\perp} \Big]\bar W + i\nu' \bar W-\lan F\ran}}
&=&
\frac{C_\perp}{4k|p|}\lt(\nabla_{\bp_\perp}-i{\beta}\bx_\perp\rt)^2
  \bar W
\eeq
where the constant $C_\perp$ is the paraxial approximation of  (\ref{diffusion}) for $|p|=1$:
\[
C_\perp=\frac{\pi}{2}\int \Phi(0,\bq_\perp)|\bq_\perp|^2
d\bq_\perp.
\]
Here we have assumed the isotropy of $\Phi$ in
the transverse dimensions. 
Note that the longitudinal (momentum) diffusion vanishes
and that  the longitudinal momentum $p$ plays 
the role of a parameter in eq. (\ref{para}) which then
can be solved in the direction of increasing $z$ as an evolution equation with
initial data given  at a fixed $z$.   
This is another instance of  paraxial approximation.

 Let $\sigma_*$ be
the spatial spread in the transverse coordinates $\bx_\perp$, $\ell_c$ the coherence length in the transverse dimensions
and $\beta_c$ the coherence bandwidth. Let $L$ be
the scale of the boundary layer.
We then seek the following change of
variables
\beq
\label{change2}
\tilde\bx_\perp=\frac{\bx_\perp}{\sigma_* k},\quad
\tilde\bp_\perp=\bp_\perp k\ell_c, \quad\tilde z=\frac{z}{Lk},
 \quad
\tilde\beta=\frac{\beta}{\beta_c}
\eeq
to remove all the physical parameters from
(\ref{para}) and to aim for
the form
\beq
\label{fp'}
\partial_{\tilde z}  \bar W+\tilde\bp_\perp\cdot\nabla_{\tilde\bx_\perp} \bar W
+ Lki\nu'\bar W-Lk\lan F\ran 
=\lt(\nabla_{\tilde\bp_\perp}-i{\tilde\beta}\tilde\bx_\perp\rt)^2\bar W.
\eeq
The same reasoning as above now leads to
\beqn
\ell_c \sigma_*\sim L/k,\quad \sigma_*/\ell_c \sim {1}/{\beta_c},\quad 
\ell_c\sim k^{-1}L^{-1/2} C_\perp^{-1/2}
\eeqn
and hence  \[
\sigma_*\sim L^{3/2} C_\perp^{1/2},\quad
\beta_c\sim k^{-1}C_\perp^{-1} L^{-2}.
\]
The layer thickness $L$ may be determined by $\ell_c\sim 1$, i.e. $L\sim k^{-2}C_\perp^{-1}$.

After the  inverse Fourier
transform eq. (\ref{fp'}) becomes  
\beq
\label{mean-eq2}
\partial_{\tilde z} \Gamma
-{i}\nabla_{\tilde\by_\perp}\cdot\nabla_{\tilde\bx_\perp} \Gamma + Lki\nu' \Gamma-Lk\lan F\ran
&=&-\big|\tilde\by_\perp+{\tilde\beta}\tilde\bx_\perp\big|^2
\Gamma\eeq
 which is the governing equation for the two-frequency
 mutual coherence in the normalized variables. With 
 data given on $\tilde{z}=0$ and vanishing far-field boundary
 condition in the transverse directions,  Eq. (\ref{mean-eq2}) can be solved analytically 
and its Green function  is
given by
\beq
\label{asym}
&&\frac{e^{-iLk\nu'}(i4\tilde\beta)^{1/2}}{(2\pi)^2\tilde z\sinh{\big[(i4\tilde\beta)^{1/2}\tilde z\big]}}  
\exp{\lt[\frac{1}{i4\tilde\beta\tilde z}
\lt|\tilde\by_\perp-\tilde\beta\tilde\bx_\perp-\by'_\perp+\tilde\beta\bx'_\perp\rt|^2\rt]}
\\
\nn&&\times \exp{\lt[{-\frac{\coth{\big[(i4\tilde\beta)^{1/2}\tilde z\big]}}{(i4\tilde\beta)^{1/2}}
\lt|
\tilde\by_\perp+\tilde\beta\tilde\bx_\perp
-\frac{\by'_\perp+\tilde\beta\bx'_\perp}{
\cosh{\big[(i4\tilde\beta)^{1/2}\tilde z\big]}}\rt|^2}\rt]}\\
\nn&&\times
 \exp{\lt[-\frac{\tanh{\big[(i4\tilde\beta)^{1/2}
\tilde z\big]}}{(i4\tilde\beta)^{1/2}}
\lt|\by'_\perp+\tilde\beta\bx'_\perp\rt|^2\rt]}.
\eeq

Formula (\ref{asym})  is consistent with
the asymptotic result  in the literature 
which mainly concerns with the cross-frequency correlation
of {\em intensity}. In the radiative transfer regime considered
here, the cross-spectral correlation of intensity is the square
of 
the two-frequency mutual coherence and has the commonly accepted form \cite{Sha, Gen, RN}
\beq
\label{current}
\exp{\Big[-2\sqrt{2\tilde\beta}\Big]}
\eeq
which is just the large $\tilde\beta$ asymptotic of the squared factor
$|\sinh{[(i4\tilde\beta)^{1/2}\tilde z]}|^{-2}$ in (\ref{asym}) 
at $\tilde z=1$ (see \cite{2f-grt} for detailed comparison).
Moreover (\ref{asym}) 
provides
 detailed information about the simultaneous
dependence of the mutual coherence on the frequency difference and spatial displacement  for $\tilde z\in (0,1)$ \cite{RN, SHG}.  

Surprisingly, a closely related equation arises in the two-frequency formulation of the Markovian
approximation of the paraxial waves \cite{2f-whn}. 
The closed form  solution is crucial for analyzing the performance
of time reversal communication with broadband signals
\cite{pulsa}. The solution procedure for (\ref{asym})
is similar to that given elsewhere \cite{pulsa} and 
is omitted here. 

\subsection{Paraxial 2f-GRT in anisotropic media}
\label{gan}
 We use here the setting and  notation defined in Section \ref{prt}  for anisotropic media. For simplicity we will set $p=1$ 
and omit writing it out in $\bar W$. 
In view of  (\ref{aniso}) we replace $\Phi(\bq)$ in (\ref{diffusion}) by
\[
\delta(q) \int dw \Phi(w,\bq_\perp)
\]
and obtain the transverse diffusion coefficient 
\[
\bD_\perp(\bp_\perp)=\pi \int d\bq_\perp \int dw \Phi(w,\bq_\perp)
\delta(\bp_\perp\cdot\bq_\perp)\bq_\perp\otimes\bq_\perp
\]
whereas the longitudinal diffusion coefficient is zero. 

For simplicity we assume the isotropy in the transverse dimensions, $\Phi(w,\bp_\perp)=\Phi(w,|\bp_\perp|)$, so that
 $\bD_\perp={C_\perp} |\bp_\perp|^{-1} \Pi_\perp(\bp_\perp)$ where 
 \[
{C_\perp}=\frac{\pi}{2}\int\delta\Big(\frac{\bp_\perp}{|\bp_\perp|}\cdot\frac{\bq_\perp}{|\bq_\perp|}\Big)\Phi(w, |\bq_\perp|)|\bq_\perp| dw d\bq_\perp
\]
is a constant 
and 
$\Pi_\perp(\bp_\perp)$ is 
the orthogonal projection onto the transverse line perpendicular to
$\bp_\perp$.
Hence eq. (\ref{go}) reduces to
\beq
\label{go-para}
\lefteqn{\Big[\partial_{z}+{\bp_\perp\cdot\nabla_{\bx_\perp} \Big]\bar W + i\nu' \bar W-\lan F\ran}}\\
&=&
\frac{C_\perp}{4k}\lt(\nabla_{\bp_\perp}-i{\beta}\bx_\perp\rt)\cdot |\bp_\perp|^{-1} \Pi_\perp(\bp_\perp)
\lt(\nabla_{\bp_\perp}-i{\beta}\bx_\perp\rt)
  \bar W. \nn
\eeq
Alternatively, eq. (\ref{go-para}) can also be derived from eq. (\ref{rt-para})
by taking the geometrical optics limit as described in the beginning
of Section 6. 

Consider the change of variables (\ref{change2}) to remove all the physical parameters from
(\ref{go-para}) and to aim for
the form
\beq
\label{go-para-std}
\lefteqn{\Big[\partial_{\tilde z}+{\tilde \bp_\perp\cdot\nabla_{\tilde \bx_\perp} \Big]\bar W +Lk i\nu' \bar W-Lk\lan F\ran}}\\
&=&
\lt(\nabla_{\tilde\bp_\perp}-i{\tilde\beta}\tilde\bx_\perp\rt)\cdot |\tilde\bp_\perp|^{-1} \Pi_\perp(\tilde\bp_\perp)
\lt(\nabla_{\tilde\bp_\perp}-i{\tilde\beta}\tilde\bx_\perp\rt)
  \bar W\nn
\eeq
where $L$ should be interpreted as the distance of propagation.

Following the same line of reasoning, we obtain that
\[
\ell_c\sigma_*\sim L/k,\quad \sigma_*/\ell_c\sim 1/\beta_c,\quad\ell_c\sim C_\perp^{-1/3} L^{-1/3} k^{-1}
\]
and hence
\[
\sigma_*\sim C_\perp^{1/3} L^{4/3},\quad
\beta_c\sim C_\perp^{-2/3} L^{-5/3}k^{-1}.
\]

Unlike (\ref{para}) it is unclear if  a closed-form solution to eq. (\ref{go-para}) exists or not.

\section{ Discussion and conclusion}

The standard (one-frequency) RT can be formally derived from the wave equation in at least two ways: the diagrammatic expansion method, as the ladder approximation of the Bethe-Salpeter equation
\cite{RN, Mis}, 
and the multi-scale expansion method advocated here \cite{BLP}. The latter is considerably simpler than the former in terms
of the amount of  calculation involved. Both approaches have been developed with full mathematical rigor
in some special cases (see \cite{rad-arma, rad-crm} and the references therein).  There are two regimes for which the 2f-RT equation has been  derived with
full mathematical rigor: first, 
for 
the paraxial wave equation 
by using the so called martingale method in probability
theory \cite{2f-crp, 2f-rt-physa}; second, for the spherical waves in geometrical optics by  the path-integration method \cite{2f-grt}. 
These rigorous results coincide with those derived here for
the respective regimes and hence support the validity 
of MSE. 

Within the framework of 2f-RT, a paraxial form arises naturally
in anisotropic media which fluctuate slowly in the longitudinal
direction. Another form of paraxial 2f-RT takes place
in the boundary layer asymptotics of isotropic media. 
The latter equation turns out to be exactly solvable
and the  boundary layer behavior is given in a closed form, revealing highly 
 non-trivial structure  of the two-frequency mutual coherence. 
In any case, dimensional analysis with the 2f-GRT  equations
yields  qualitative scaling  behavior
of the spatial spread, the spatial frequency spread
and the coherent bandwidth in various regimes.

\commentout{
On the other hand,  
the two-frequency radiative transfer limit for  (\ref{helm}) may be dealt with  
by extending the diagrammatic method 
developed for the time dependent Schr\"odinger
equation  \cite{Sp, Sp2, HLW, EY}. 
By analogy, (\ref{helm})
is like  the stationary Schr\"odinger equation with an energy-dependent potential.  
However, the diagrammatic approach, rigorous or not,  is more complicated than MSE 
to carry out, 
so, in addition to the formal expansion, we will be content  to give a brief  analysis of  MSE in Appendix B. 
}

\commentout{
Using  MSE we have given a formal derivation of  the two-frequency radiative transfer equation for the classical wave equation in
terms of the new two-frequency Wigner distribution.  
The validity of the derivation is supported by
the interchangeability of the paraxial approximation
and the two-frequency radiative transfer limit. 
}

 
 From the point of view of computation, especially Monte Carlo 
 simulation, it appears to be natural to introduce the new
 quantity
 \[
\fW(\bx,\bp)=e^{-i\beta\bx\cdot\bp} \bar W(\bx,\bp)
 \]
 and rewrite eq. (\ref{wb}) in the following form
 \beqn
\label{wbtilde}
\lefteqn{\bp\cdot\nabla_\bx  \fW+i\beta|\bp|^2 \fW+ i\nu' \fW-e^{-i\beta\bx\cdot\bp}\lan F\ran}\\
&=&\frac{\pi k^{3}}{\theta^{4}}
\int  d\bq \Phi\big(\frac{k}{\theta}(\bp-\bq)\big)\delta(|\bp|^2
-{|\bq|^2})\Big[
\fW \big(\bx,\bq\big)
-\fW (\bx,\bp)\Big].\nn
\eeqn
The solution $\fW$ can then be expressed as
a path integration over the Markov  process generated by
the operator  $\cA$ defined  by
\[
\cA \fW=-\bp\cdot\nabla_\bx \fW+\frac{\pi k^{3}}{\theta^{4}}
\int  d\bq \Phi\big(\frac{k}{\theta}(\bp-\bq)\big)\delta(|\bp|^2
-{|\bq|^2})\Big[
 \fW \big(\bx,\bq\big)
- \fW (\bx,\bp)\Big] 
\]
when $V$ is real-valued and $\Phi$ is nonnegative. 
We will pursue  this observation in a separate publication \cite{2f-grt}. 

\commentout{
The present approach can be generalized
to the polarized waves described by the Maxwell
equations or the vector wave equation, which will be 
presented 
elsewhere. 
}

 \appendix
 \section{Derivation of eq. (\ref{raw})}
 Applying the operator $\bp\cdot\nabla$ to the definition
 (\ref{0.11}) we obtain
 \beq
\nn
\bp\cdot\nabla_\bx W&=&\frac{1}{(2\pi)^3}
\int e^{-i\bp\cdot\by}2\bp\cdot\nabla_\by U_1 (\frac{\bx}{k_1}+
\frac{\by}{2k_1}){U^*_2(\frac{\bx}{k_2}
-\frac{\by}{2k_2})}d\by\\
&&
-\frac{1}{(2\pi)^3}
\int e^{-i\bp\cdot\by} U_1 (\frac{\bx}{k_1}+
\frac{\by}{2k_1}){2\bp\cdot\nabla_\by U^*_2(\frac{\bx}{k_2}
-\frac{\by}{2k_2})}d\by\nn\\
&=&\frac{2i}{(2\pi)^3}
\int \Big(\nabla_\by e^{-i\bp\cdot\by}\Big) \cdot\nabla_\by U_1 (\frac{\bx}{k_1}+
\frac{\by}{2k_1}){U^*_2(\frac{\bx}{k_2}
-\frac{\by}{2k_2})}d\by\nn\\
&&-
\frac{2i}{(2\pi)^3}
\int \Big(\nabla_\by e^{-i\bp\cdot\by} \Big) U_1 (\frac{\bx}{k_1}+
\frac{\by}{2k_1})\cdot\nabla_\by {U^*_2(\frac{\bx}{k_2}
-\frac{\by}{2k_2})}d\by.\nn
\eeq
Integrating by parts with the first $\nabla_\by$ in the above
integrals, we have
\beq
\bp\cdot\nabla_\bx W&=&-\frac{2i}{(2\pi)^3}
\int  e^{-i\bp\cdot\by}\nabla^2_\by U_1 (\frac{\bx}{k_1}+
\frac{\by}{2k_1}){U^*_2(\frac{\bx}{k_2}
-\frac{\by}{2k_2})}d\by\nn\\
&&+
\frac{2i}{(2\pi)^3}
\int  e^{-i\bp\cdot\by} U_1 (\frac{\bx}{k_1}+
\frac{\by}{2k_1})\nabla^2_\by {U^*_2(\frac{\bx}{k_2}
-\frac{\by}{2k_2})}d\by\label{a.1}
\eeq
where the other resulting terms are canceled with each other.
From  eq. (\ref{helm}), 
\beq
\label{a.2}
\nabla^2_\by U_j (\frac{\bx}{k_j}+
\frac{\by}{2k_j})=-\frac{1}{4}\Big(\nu_j+V_j(\frac{\bx}{k_j}+
\frac{\by}{2k_j})\Big)U_j(\frac{\bx}{k_j}+
\frac{\by}{2k_j})+\frac{1}{4}f_j(\frac{\bx}{k_j}+
\frac{\by}{2k_j}).
\eeq
Using (\ref{a.2}) in (\ref{a.1}) we arrive at eq. (\ref{raw}).

 \section{Weak convergence of corrector}

First we show that the corrector does not vanish in
the in the mean-square norm in any
dimension, i.e.
 $\lim_{\ep\to 0}\ep\int\lan|W^\ep_1|^2\ran d\bx d\bp>0$ 
 in general. For simplicity, consider only the term  involving,
say, $\hat V_1$ in the expression (\ref{w1'}).
A straightforward  calculation shows
  \bean
\lefteqn{\lim_{\ep\to 0}\frac{1}{4\theta^2}\int\,d\bp\,d\bx d\bq \Phi_1(\bq) {\ep\over
\ep^2+(\bp\cdot\bq/k_1)^2}
\big|\bar W(\bx,\bp-\frac{\theta\bq}{2k_1})\big|^2}\label{est}\\
&=&\frac{k\pi}{4\theta^2}
\int\,d\bp\,d\bx d\bq \Phi(\bq)\delta(\bp\cdot\bq)\big|\bar W(\bx,\bp-\frac{\theta\bq}{2k})\big|^2
\eean
which is positive  in general.

Next we show that 
the corrector  vanishes in the weak topology
\be
\label{corr2}
\lim_{\ep\to
0}\ep \int\,d\bx\,\lan\left|\int\,d\bp\,W^\ep_1(\bx,\frac{\bx}{\ep},\bp)\psi(\bp)\right|^2\ran
=0,\quad \forall \psi\in L^2.
\ee
It suffices to prove (\ref{corr2}) for any smooth, compactly supported function $\psi$. For the term involving $\hat V_1$ only, we have 
\beqn
&&\lim_{\ep\to 0}\frac{\ep}{4\theta^2}\int\,d\bp d\bp'\,d\bx d\bq {\Phi_1(\bq) \psi(\bp)\psi^*(\bp')\over
(\ep+i\bp\cdot\bq/k_1)(\ep-i\bp'\cdot\bq/k_1)}
\bar W(\bx,\bp-\frac{\theta\bq}{2k_1})\bar W^*
(\bx,\bp'-\frac{\theta\bq}{2k_1})\\
&=&\lim_{\ep\to 0}
\frac{\ep k_1^2}{4\theta^2}\int d\bq {\Phi_1(\bq) \over |\bq|^2}
\Big[\pi\int d\bp \delta(\bp\cdot \hat\bq)\psi(\bp)
\bar W(\bx,\bp-\frac{\theta\bq}{2k_1}) -
\int\cpv d\bp \frac{i\psi(\bp)}{\bp \cdot\hat\bq}
\bar W(\bx,\bp-\frac{\theta\bq}{2k_1})\Big]\\
&&\times\Big[\pi\int d\bp' \delta(\bp'\cdot \hat\bq)\psi^*(\bp')
\bar W^*(\bx,\bp'-\frac{\theta\bq}{2k_1}) +
\int\cpv d\bp'\frac{i\psi^*(\bp')}{\bp' \cdot\hat\bq}
\bar W^*(\bx,\bp'-\frac{\theta\bq}{2k_1})\Big]
\eeqn
where $\hat\bq=\bq/|\bq|$ for sufficiently smooth $\bar W, \Phi$ and rapidly decaying $\Phi$. The essential point
now is that $|\bq|^{-2}$ is an integrable singularity 
in three dimensions and hence the above expression
vanishes in the limit. 

\commentout{
In summary, the multi-scale expansion (\ref{mse}) is
to be understood in the sense of (\ref{corr2}), i.e.
strongly in $\bx$ but weakly in $\bp$ 
 in the mean-square sense.  
This is  physically reasonable as $\bp$ corresponds
to the scale of fluctuation compatible
with that of  the refractive index. 
}

\commentout{
 
 \section{ Corrector estimate}

In the linear case, Spohn \cite{Sp} has proved local-in-time convergence
of the  solution of the linear Schr\"{o}dinger equation to that
of the linear transport equation. Erd\"{o}s and Yau \cite{EY}
have improved
the result so  that the convergence is global in time.
Both results involve detailed analysis of graphs corresponding to
multiple scattering.

\commentout{
We will not strive for optimal conditions on $\hat R$ and $W$, but
rather simplicity of the argument. For example,
we assume  throughout this section that
$\hat R(\bk)$ and
$W(t,\bx,\bk)$ have compactly supports in $\bk$,
while, in fact, some mild decay as $|\bk|\to \infty$
of suffices. Note that the property
of having compact support in $\bk$ is preserved by the transport
equation (\ref{2.3.7}).
We also assume that $\hat R$ is continuous. The regularity
on $W$ will be specified separately in each of the following propositions.
}

As pointed out in Section \ref{RID} , the initial data for the Wigner
function does not converge strongly, so neither the solution $W^\eps$
of the Wigner equation
nor the
corrector $W_1$ is expected to converge strongly.
This is, indeed, the case as stated next.
\begin{proposition}
\label{P1}
Suppose $\bar W(\bx,\bp)$ is such that the function
\[
f(\bq,\bp)=\Phi(\bp) \int\,d\bx\,
\left[\bar W(\bx,\bq-\bp/2)-\bar W(\bx,\bq+\bp/2)\right]^2
\]
is continuous and its zero set does not contain the set
$\{(\bq,\bp): \bq\cdot\bp=0\}$.
Then, $\ep\int\langle|W_1|^2\rangle\,d\bx d\bp$ does not vanish
as $\ep\to 0$, in any dimension.
\end{proposition}

{\em Proof.} A straightforward calculation  leads to
\bea
\nonumber
\hspace{1cm}\lefteqn{\ep\int\int\langle|W^{(1)}_\ep|^2\rangle\,\,d\bx d\bk}\\
\nonumber
\hspace{1cm}&=&\int\,d\bp\,\Phi(\bp)\int\,d\bk\,{\ep\over
\ep^2+(\bk\cdot\bp)^2}\int\,d\bx\,
\left[W(\bx,\bk-\bp/2)-W(\bx,\bk+\bp/2)\right]^2\\
&\geq&{1\over 2\ep}
\int\,d\bk\, \int_{|\bk\cdot\bp|\leq\ep}\,d\bp\,\Phi(\bp)
\int\,d\bx\,\left[W(\bx,\bk-\bp/2)-W(\bx,\bk+\bp/2)\right]^2
\nonumber\\
&\geq&
{1\over 2\ep}\int\,d\bk\,\int_{|\bk\cdot\bp|\leq\ep}\,d\bp\,f(\bk,\bp)
\label{a.1}
\eea
As the set
$\{\bk\in R^d: |\bk\cdot\bp|\leq \ep\}\,\,\bigcap \,\,
\hbox{supp}\{f(\bk,\bp)\}
$
has a measure
of order $\ep$ for $W$ satisfying the stated assumption,
the expression (\ref{a.1}) does not vanish 
in the limit $\ep\to 0$, as we wanted to show.  

We note that ``generic'' functions $W$
that are {\em not} spherically symmetric
in $\bk$ satisfy the condition stated in Proposition~\ref{P1}.

However, $\sqrt{\ep}W_1$ does vanish strongly in $\bx$
if it is first integrated against a test function  $\psi(\bk)$, as
stated in the next proposition.

\begin{proposition}
\label{P2}
For $d\geq 3$ and any differentiable $W(\bx,\bp)$ with a compact
support, we have
\be
\label{a.2}
\lim_{\ep\to
0}\ep \int\,d\bx\,\lan\left|\int\,d\bp\,W_1\phi(\bp)\right|^2\ran
=0,\quad \forall \phi(\bp)\in C^\infty_c
\ee
\end{proposition}
{\em Proof.}
After taking the expectation,
the expression on the left side of (\ref{a.2}) becomes
\be
\ep\int\,d\bp\,{\Phi(\bp)\over |\bp|^2}\int\,d\bx\,\left|
\int\,d\bk\,{\phi(\bk)\over \ep-i\bk\cdot\hat\bp}
\left[W(\bx,\bk-\bp/2)-W(\bx,\bk+\bp/2)\right]\right|^2
\label{a.4}
\ee
where $\hat\bp=\bp/|\bp|$. Since $|\bp|^{-2}$ in (\ref{a.4})
is an integrable singularity in three or more dimensions,
the expression in (\ref{a.2}) is $O(\eps)$ as $\eps\to 0$,
as we wanted to show.
}


\end{document}